\documentclass[a4paper,12pt]{article}
\usepackage{graphics}
\usepackage{multicol}
\usepackage{fullpage}
\usepackage{geometry}
\usepackage{amsfonts}
\usepackage[usenames,dvipsnames,svgnames,table]{xcolor}
\begin{document}
\title{Optical coherence in astrophysics :\\ The powerful alternative of big bang.}
\author{Jacques Moret-Bailly.\\email:jmo@laposte.net}

\maketitle

\abstract{
The coherence of the interaction of light with a collisionless gas (Einstein 1917) founds the theory of gas lasers. It is, for the understanding of universe, a simpler and more powerful tool than the big bang which requires questionable supplements (dark matter, MOND, etc.. ). The Impulsive Stimulated Raman Scattering (ISRS) redshifts gradually light pulses which cross excited atomic hydrogen H*, so that the redshift is a measure of the column density of H*. Thus, the distance of the hot stars, surrounded by much H*, is exaggerated by the use of Hubble's law. Local exaggerated distances create voids in the maps of galaxies which become spongy. The interpretation of spectra of quasars, the periodicity of galaxy redshifts introduce an experimental ``Karlsson's constant" exactly computed by ISRS. The need for dark matter comes from the exaggeration of the distance, therefore the size of galaxies. Without dark matter, celestial mechanics provides a reliable distance of spiral galaxies. Coherence also introduces superradiance and mode competition that explain that only the limbs of Str\"omgren spheres are visible as circles maybe punctuated by an even number of dots: Too numerous, the figures assigned to gravitational lenses can be such limbs. The coincidence of the ignition of the rings of SNR1987A with the extinction of the star is due to a multiphoton coherent scattering of star light, which amplifies the superradiant emission of the rings.. A blueshift of microwaves crossing H* resulting, between 10 and 15 AU, of the expansion of solar wind, explains the "anomalous acceleration" of Pioneer probes. All is obtained without any change in theories of standard spectroscopy.}


    52.35.Mw: Nonlinear phenomena: waves, wave propagation, and other interactions.

    52.38.Bv: Rayleigh scattering; stimulated Brillouin and Raman scattering.

    52.40.Db. Electromagnetic (nonlaser) radiation interactions with plasma.

\begin{multicols}{2}

\section{Introduction} \label{intro}
In a paper in Reviews of Modern Physics entitled \textquotedblleft Laser physics: Quantum controversy in action.''  \cite {WLamb2} W. E. Lamb Jr, W. P. Schleich, M. O. Scully and C. H. Townes criticize an abusive use of the word photon. For them, in accordance with the principles of quantum electrodynamics, the photon is a pseudo-particle resulting from the quantization of normal modes of the electromagnetic field. These modes should not only be defined, but also set by a physical system such as a cavity or a pulse propagating in a diffraction limited beam. 

A diffraction limited beam may be obtained putting a diaphragm on a wave surface of a monochromatic plane wave. The diaphragm is the waist of the beam which, including the virtual side, expands in the past and the future. As the transverse extension of a pulse is very large at a long distance, the field defining the mode may excite atoms in a large volume. In quantum theory, a large set of identical atoms may be considered as a single, highly degenerated system which interacts with the pulse, then, by decoherence, transfers a quantum to a single atom. In classical theory, all atoms are slightly excited, then decoherence results from an electromagnetic transfer of energy to an atom favored by a fluctuation of the zero point field. 

\medskip 
On the contrary, some physicists believe that the photon is a small, real particle:

In 1931, Menzel \cite{Menzel} wrote: {\it It is easily proved that the so-called ``stimulated emissions'' are unimportant in the nebulae}.  Menzel's demonstration is doubtful because in this paper, he confuses two photometric quantities, radiance and irradiance: The radiance measures the energy flow in a light beam, it is constant in a transparent medium, while the irradiance is the flow, coming from all directions, which crosses from a given side, a surface of unit area. This confusion was common at this time because these two quantities differ by a solid angle often not taken into account in the equations of dimension. Menzel's opinion remains in astrophysics: For most astrophysicists, the small photon in the extremely low pressure gas of a nebula can interact only with a single molecule. This interaction has a stochastic nature requiring, for instance the use of a Monte-Carlo draw \cite{Zheng}.

 Without taking sides in this controversy, it is useful to precise which theory of the propagation of light is used. Here, we follow the choice done by Lamb et al. : The interaction of an electromagnetic wave with a molecule follows to a deterministic process, so that Huygens' construction may apply and the interaction is coherent. 

The view of  Lamb et al. is supported experimentally by the absence of blue sky in the stratosphere and the need to avoid incoherent diffusions in gas lasers: {\it  All light interactions with low pressure gas are coherent.} Incoherences generating the blue sky, result of collisions, mostly binary, whose density is proportional to the square of the pressure.

\medskip
 To simplify, in our models, nebulae consist of pure hydrogen.

\medskip

Section 2 summarizes the general properties of coherent spectroscopy.

Section 3 studies, using coherent spectroscopy, theoretical models involving extremely hot sources, huge systems, which may exist in astronomy.

Section 4 compares some observations with these models.

\section{Coherent interactions.}
\subsection{Huygens' construction.}

\begin{figure*}
\begin{center}
\resizebox{\columnwidth}{!}{\includegraphics{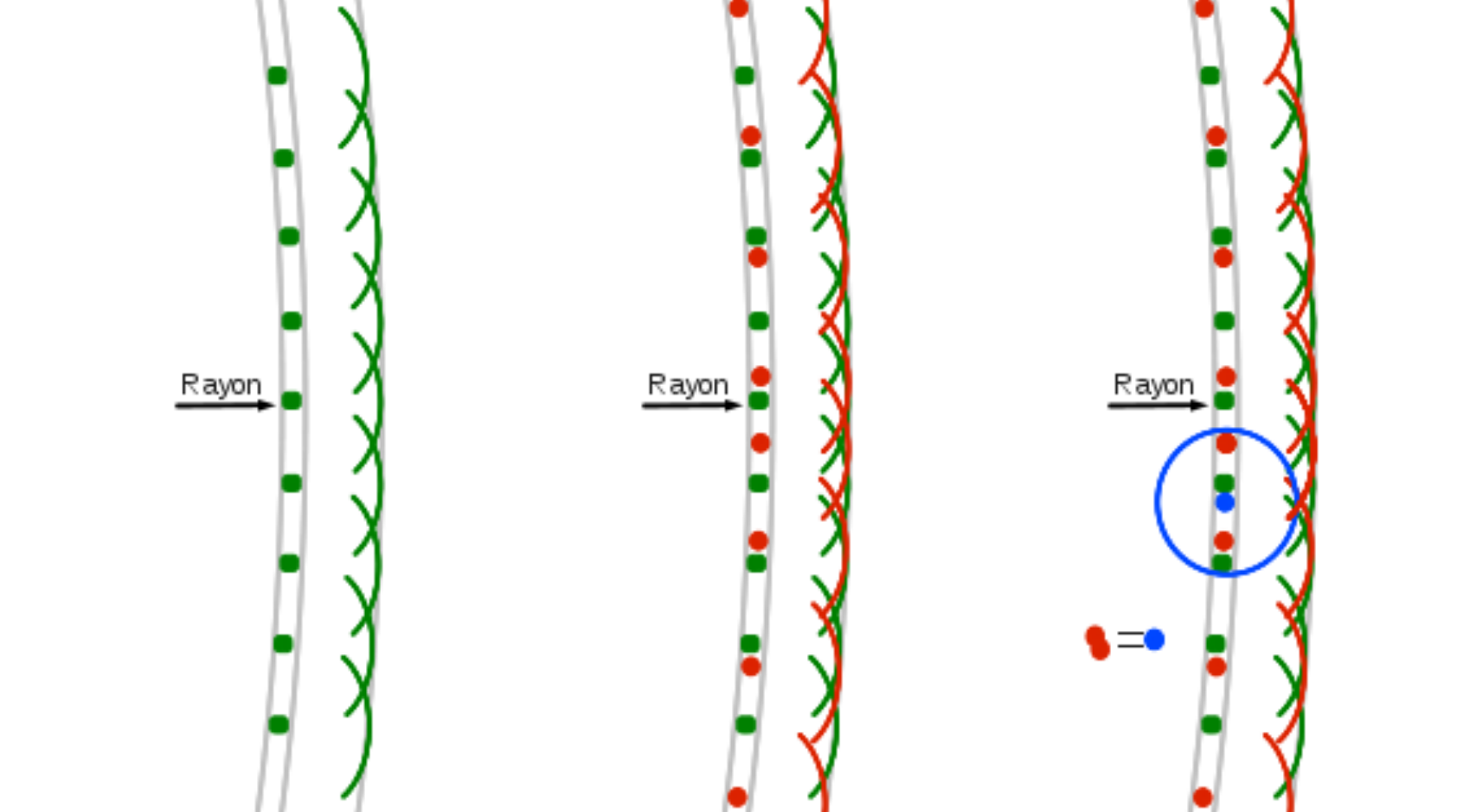}}
\end{center}
\caption{\label{rad} A, B, C.--Huygens' constructions: A in a \textcolor{ForestGreen}{continuous medium}; B in a \textcolor{red}{low pressure gas}; C \textcolor{blue}{including rare molecules} possibly made up by two colliding molecules.}
\end{figure*}

Huygens' constructions for any sine waves are representations of the generation of a new wave surface near an old wave surface: Sources located in the vicinity of the old surface wave radiate synchronous spherical waves.
The envelope of the spheres obtained after a time $\Delta t$ is a new wavefront.
The part of the envelope corresponding to a backward wave is eliminated because a wavefront having an initial offset of a quarter wavelength, generates back a wave of opposite phase which destroys the backward wave.

Figure 1A is Huygens' construction in an homogeneous medium. The volume between two wave surfaces whose distance is infinitesimal is divided into infinitesimal parcels which radiate in phase to build the new wavefront.

Figure 1B, on the diffusion of an electromagnetic wave, adds, to diffusion 1A, a diffusion by a large number of molecules. In a transparent medium, this last release is delayed by $\pi/2$, so that the sum of two sine waves is a sine wave slightly delayed, called refracted.

Figure 1C shows that the wavelets  emitted by rare molecules cannot generate an envelope: it is an incoherent scattering. In gases, the collisions of two molecules create binary systems depending on many parameters, so it creates different molecular systems that produce incoherent diffusions. 

Similarly, a gas laser can only operate if the incoherent diffusions are quite unlikely. To avoid a loss of too much energy, the pressure  is about 100 Pa.

\subsection{Superradiance.}

When a light beam propagates in a gas, its spectral radiance $I(\nu)$ varies according to the law given by Einstein in 1917 \cite{Einstein}
\begin{equation}
{\rm d}I(\nu)=B(\nu)I(\nu) {\rm d}x \label{ein}
\end{equation}
where d$x$ is the  path in the gas and $B(\nu)$ Einstein's coefficient which depends on the nature and state of the gas. (We generally imply ``$(\nu)$'' and ``spectral'', which specify that it is a flow of energy within a range of frequencies d$\nu$ or wavelengths of d$\lambda$).
The integration of equation \ref{ein} introduces a constant we write $I_0$ assuming it is the absolute radiance of a ray from a blackbody at 0 K, so that $I_0 = h\nu^3/c^2$  on average \footnote{Including both polarized modes.} (Planck 1911 \cite{Planck}). If $B$ is constant, $I$ is an exponential function of $x$ path.
It is often appropriate to say that there is a superradiance when $I$ is greater than $ 2I_0 $, while $I$ usually  remains close to $ I_0$.

\subsection{Competition of modes.}
\begin{figure*}[]

\begin{center}
\resizebox{6cm}{!}{\includegraphics{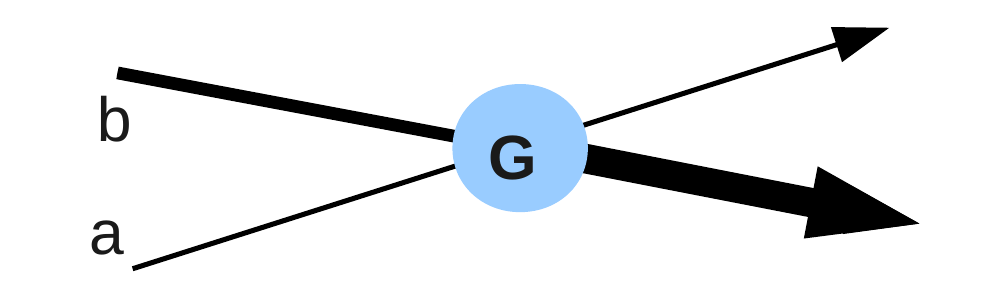}}
\end{center}
\caption{Competition of modes.}
\label{compet}
\end{figure*}

Amplified in a volume G, (fig \ref{compet}) the rays absorb energy, so that coefficient $B$ decreases. Formula \ref{ein} shows that the strongest ray absorbs more energy than the weaker, the radiance of which remains close to $I_0$: It remains bright only the rays which would be the strongest taking no account of superradiance.

\subsection{Multiphotonic interactions.}

Suppose that an extremely hot source emits a continuous spectrum. At each frequency, its high radiance beams can pump atoms to virtual levels, from which they may be pumped, and other pumpings may allow them reach a stationary high level: by combinations of frequencies, the whole continuous spectrum is involved in the pumping of the atoms to stationary levels.

If the decay from this stationary level feeds a superradiant beam, the $B$ coefficient may remain high, so that there is a transfer of energy from the whole continuous spectrum of the source to the superradiant beam.

This transfer is only possible if it increases  the entropy of the system. The temperature of a light ray at a frequency $\nu$ is deduced by Planck's law from its spectral radiance and $\nu$. To increase the entropy, the temperature of the rays emitted by the source and the temperature of the superradiant beam tend to a common limit: A  fraction of the energy of the continuous spectrum emitted by the source is transferred to the superradiant beam.

\subsection{Impulsive Stimulated Raman Scattering (ISRS)}\label{isrs}
This scattering is widely used in Keith Nelson's laboratory in MIT, mainly to study matter. A simple description may be found in paper \cite{Yan}, for a review of ISRS see, for instance, Dhar et al. \cite{Dhar} or Lee et al. \cite{Lee}.

The ISRS may be considered as a supplement of refraction obtained by an addition of a Raman coherent scattering to the Rayleigh coherent scattering which produces the refraction (Look at appendix). Just as the Rayleigh coefficient of scattering, the Raman coefficient of scattering deduced from the quadrupolar polarizability is, in first approximation, far from the resonances, proportional to the frequency of the incident, exciting ray.

However, in refraction, the incident and Rayleigh scattered beams have the same frequency, so that they add into a wave at the same frequency. In ISRS, the exciting and coherently scattered beams have different frequencies, beats appear.

The fields scattered at two points A and B of a ray whose distance is $x$, have the same phase-shifts versus the incident wave. Along the path AB, they get the phase-shift  $\Delta \phi =2\pi x(1/\lambda_i-1/\lambda_d)$, where $\lambda_i$ and $\lambda_d$ are the wavelengths
of the incident and scattered waves. $\lambda_i$ and $\lambda_d$ differ except for particular rays propagating in some crystals. Thus, assuming a negligible variation of the intensity of the incident beam, for a distance $x$ such that $\Delta \phi$ equals $\pi$, the scattered amplitudes cancel.

\begin{figure*}[]

\begin{center}
\resizebox{\columnwidth}{!}{\includegraphics{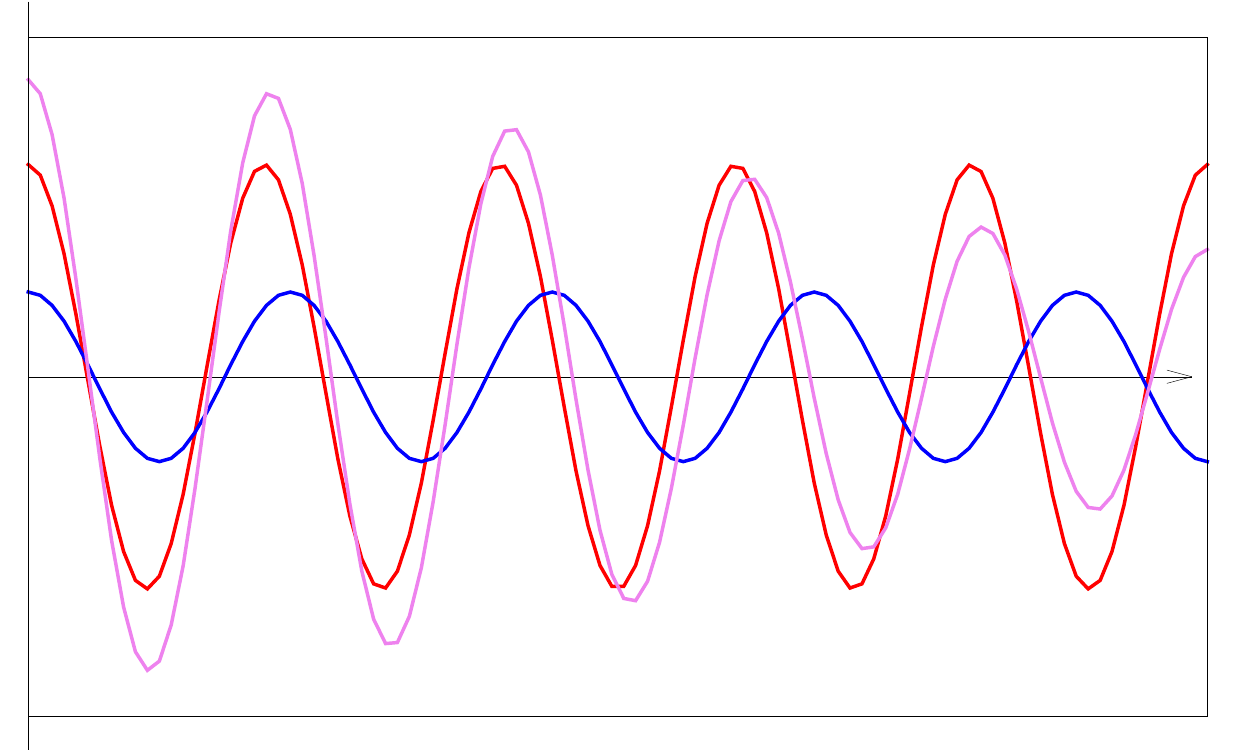}}
\end{center}
\caption{Addition of sine wave of different frequencies: While two sine curves \textcolor{red}{\bf *} \textcolor{blue}{*} have a constant amplitude, their sum\textcolor{violet}{*} shows the beginning of a beat.}
\label{Addsin}
\end{figure*}

Figure \ref{Addsin} shows, with unrealistic orders of magnitude, interference of an incident wave with an other whose frequency is slightly different and amplitude  lower,  playing the role of a Raman Stokes wave. The amplitude of the sum of both waves decays until their phases are opposed. This sum has the appearance of a sine wave whose frequency decreases then increases.

The decay zone is selected by stopping the interaction by the use of pulses ``shorter than all involved time constant'' (G. L. Lamb \cite{GLamb} ). If this condition is met, the beating of incident and scattered waves does not appear and their sum has a sinusoidal appearance. Fourier analysis shows that the major frequency is intermediate between initial frequencies in the ratio of the amplitudes of the components. 

The computation of the frequency shift in the appendix shows that :.

- The properties of the ISRS that do not involve temporal variations throughout the duration of the pulse are obtained by a simple transposition of the properties of refraction:

- Refraction and ISRS are spatially coherent, allowing the formation of images.

- Index of refraction, and relative variation of frequency by ISRS $\Delta f/f$ depend on a dispersion which is low except in the vicinity of resonances.

The ISRS frequency shift is easily observed by using femtosecond laser pulses \cite{Yan}.

\medskip

The ISRS excites the active medium during the pulses. The decay of this excitation can be probed by another ISRS. It can also be probed with a continuous laser beam which undergoes a frequency variation during the pulsed excitation of the Raman level.

\medskip
A coupling of several ISRS so that the balance of excitations of the medium equals  zero is called `` parametric''.  The matter is a catalyst and the frequency shifts are much more intense because the Raman transitions are not saturated. Evidently, energy exchanges must be in accordance with the principles of thermodynamics.

\subsection{Coherent Raman Effect on Incoherent Light (CREIL).}
CREIL is an ISRS wherein the pulses result from the usual representation of the temporally incoherent light emitted by thermal sources. The 10 femtosecond pulses are replaced by nanosecond pulses. The appendix shows that an increase of the length of the pulses by a factor $k$ decreases the order of magnitude of ISRS by a factor  $k^3 = 10^{15}$, so that an observation of CREIL requires an astronomical path of interaction.
 
A thermal electromagnetic field is always present, always tending to lower the frequency of the light.

{\it The only gas, abundant in the universe, having a resonance period of more than, and close to 1 ns, thus able to catalyze a CREIL, is excited atomic hydrogen in the levels n = 2}, denoted H*: it has quadrupole transitions at 178 MHz in the state 2S$_{1/2}$, 59 Mhz in  2P$_{1/2}$ and 24 MHz in 2P$_{3/2}$. More excited levels have resonances of lower frequencies, their effect being almost negligible.
 
\section{Astronomical models.}
\subsection{Str\"omgren's sphere \cite{Stromgren}.}
\begin{figure*}
\begin{center}
\resizebox{5cm}{!}{\includegraphics{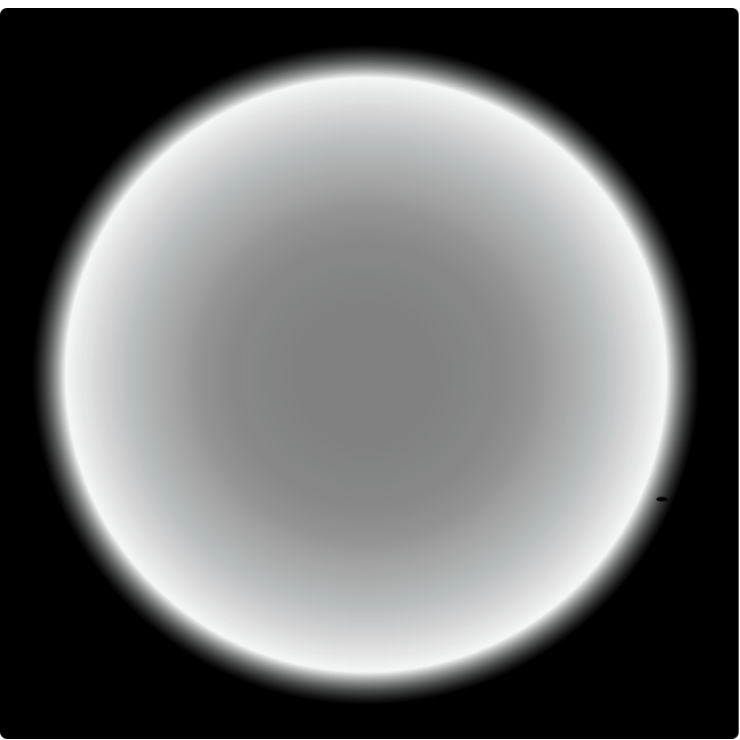}}
\end{center}
\caption{Str\"omgren's shell without superradiance.}
\label{sphere}
\end{figure*}

In 1939, Str\"omgren \cite{Stromgren} has defined a system consisting of a source (star) very hot (several hundreds of thousands of kelvins) placed in a cloud of hydrogen, at low pressure, initially cold.

The source generates a sphere of plasma almost completely ionized into protons and electrons. Str\"omgren showed that some atoms created by the cooling of the plasma radiate much energy, which accelerates the cooling and creates new atoms in a catastrophic process. It therefore appears a shell of plasma containing hydrogen atoms that radiate their spectral lines intensely.
Str\"omgren seems to have ignored the superradiance, so that his system had the appearance of figure \ref{sphere}.

\begin{figure*}
\begin{center}
\resizebox{\columnwidth}{!}{\includegraphics{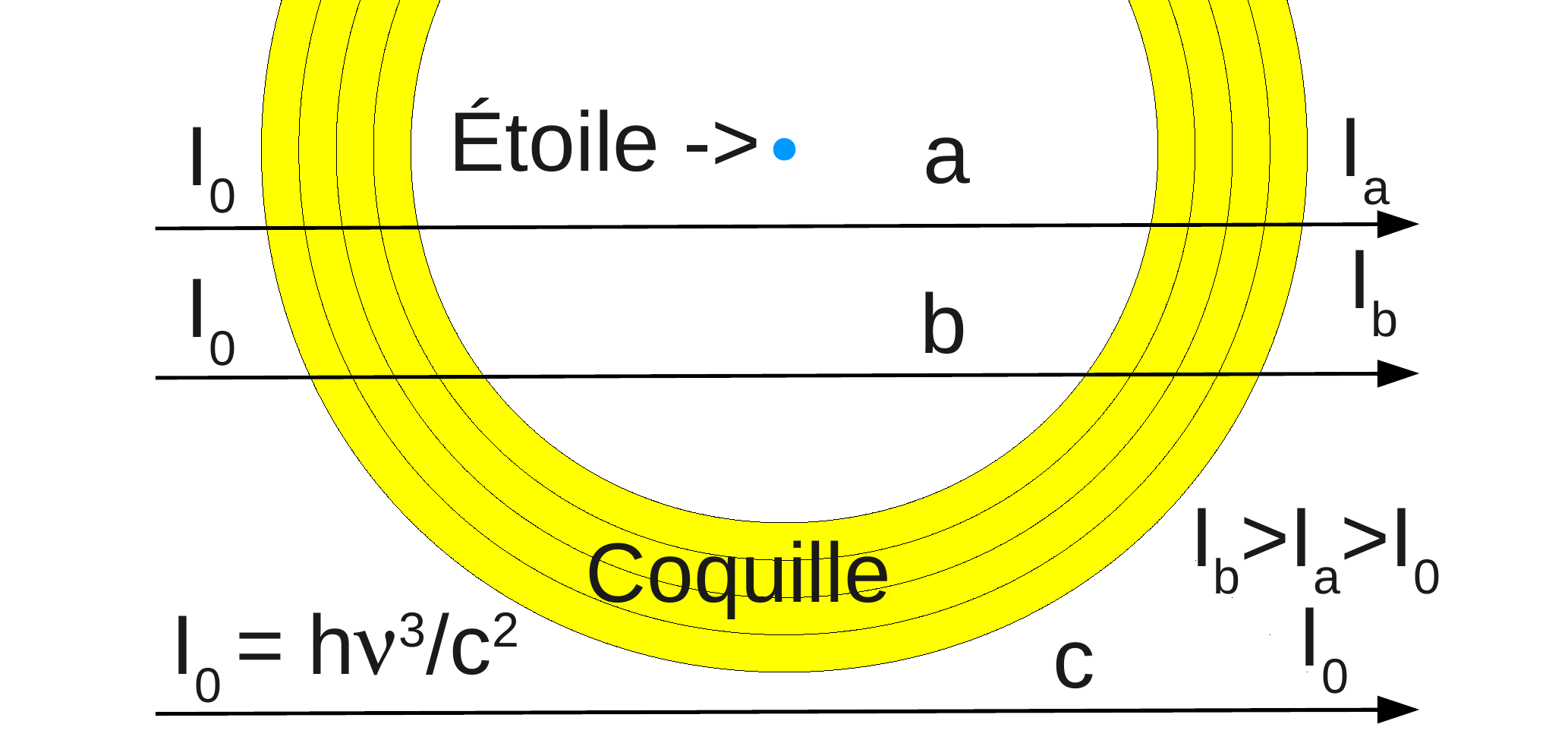}}
\end{center}
\caption{Emission of the shell (coquille).}
\label{coquille}
\end{figure*}

The shell is divided into an infinite number of sub-layers (Fig. \ref{coquille}). As a ray intersects the sphere, the path in each sub-layer is an increasing function of the distance $r$ of the ray to the star. So radiance $I$ is a function of $r$, first increasing , finally falling to 0, so it has at least a maximum $R$ which specifies the radius of the Str\"omgren's sphere. Accordingly, only the rays at the limb of the sphere are brilliant.

\begin{figure*}

\begin{center}
\resizebox{5cm}{!}{\includegraphics{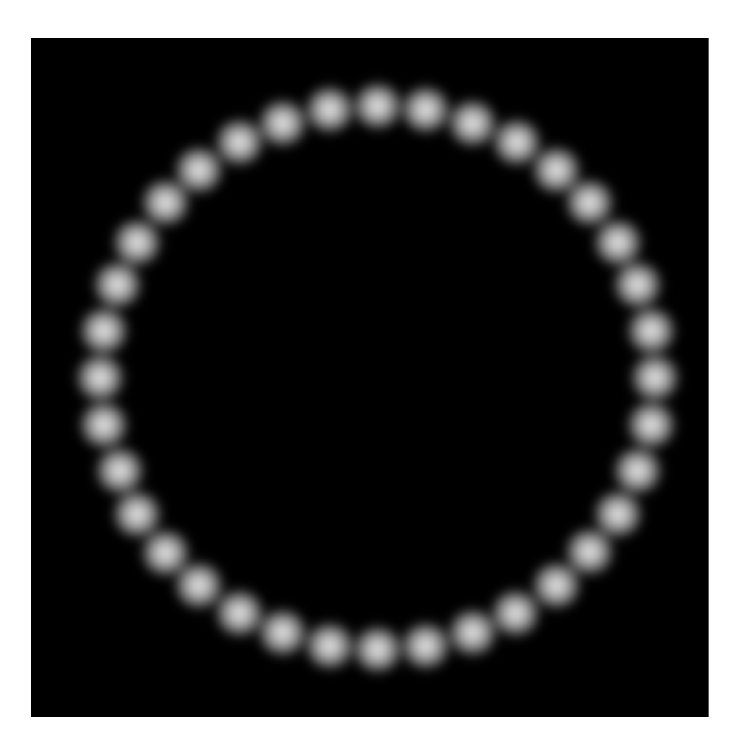}}
\end{center}
\caption{Modes of Str\"omgren's shell.}
\label{modes}
\end{figure*}

If superradiance is sufficient the symmetry is broken by competition of modes in the limb. The limb is seen as an even  number of points  because the phases of two adjacent modes are opposite (Fig. \ref{modes})

\begin{figure*}
\begin{center}
\resizebox{\columnwidth}{!}{\includegraphics{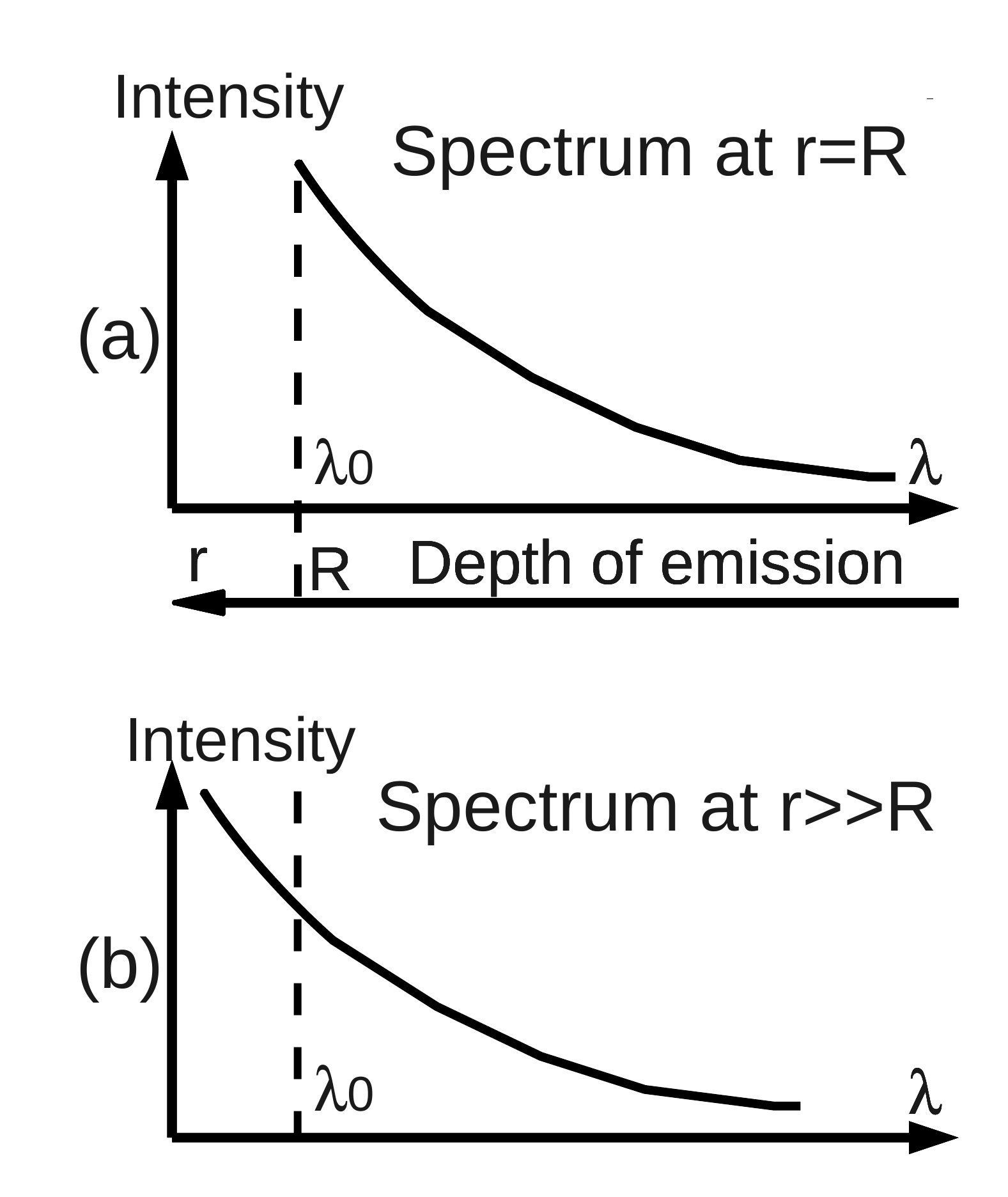}}
\end{center}
\caption{Spontaneous emission of Ly$\alpha$ line on a ray observed inside the main ring of SNR1987A.}
\label{specly}
\end{figure*}

\subsection{Emission inside a Str\"omgren's sphere.}
The sphere contains mostly free protons and electrons that do not interact with light. A low density of excited atoms increases nearly exponentially in terms of $r$, to the surface.

Suppose impurities absorb enough frequencies around Lyman alpha inside the sphere, so that the radiance of the rays which arrive from inside in vicinity of the surface is $h\nu^3/c^2$. Study light in the vicinity of the exit point of the sphere :

- In depth, the emission (amplification) of a line, for example Ly $\alpha$ at $\lambda_0$ is small, then it is redshifted by CREIL effect along the path in gas to the surface.

- Approaching the surface, the emission grows exponentially, while the redshift decreases. (Figure \ref{specly} (a)).

- Above the surface, the light passes through a region in which the irradiance of superradiant UV is large, so that the frequencies of the low radiance spectrum are increased (Figure \ref{specly} (b)). In terms of increasing wavelength, the emission line appears suddenly at a wavelength $\lambda_1 <\lambda_0$, then decreases exponentially . In practice, the observation adds distinct spectra of rays, so that the intensity growth is less abrupt.

\subsection{Absorption of a continuous UV-X spectrum by relatively cold (30 000K) atomic hydrogen.}\label{peri}

Suppose that a star is surrounded by hydrogen whose temperature and pressure decrease with altitude. Follow the light emitted by the star.

The star emits a high frequency continuous spectrum. Close to the star, hot, excited atomic hydrogen emits the Lyman lines which are much broadened by the pressure.

The excited states are nearly depopulated by a decrease of temperature under 50 000 K. The absorption  at Lyman frequencies prints the spectrum in light. The pressure drop sharpens the absorbed spectral lines. The absorption of Lyman $\alpha$ line populates the 2P states of the atoms, so that, where the duration of the free path of the atoms becomes larger than a nanosecond, a CREIL appears, which redshifts continuously the frequencies of light.The fast passing of the absorbing lines through the spectrum absorbs little light, so that the broad, weakly absorbed lines are invisible.

However, if an absorbed line reaches by redshift the Lyman $\alpha$ frequency, 2P atoms are not created, the redshift stops, so that the whole absorption spectrum of H is written clearly in the light.

But the stop is not perfect because excited $n>2$ states provide a weak redshift and may decay into $n=2$ states: the redshift restarts.

 Thus, two Lyman spectra are written in the light, one at its regular frequencies, the other with the redshift $Z_{(\beta,\alpha)} = (\nu_\beta-\nu_\alpha)/\nu_\alpha$ which puts the Lyman $\beta$ line to the $\alpha$. The process may restart, in which the lines of both absorbed spectra may generate the absorption of a new spectrum.

The final result is a set of absorption lines in which all lines result from the absorption of Lyman spectra redshifted by successive redshifts which shift various Lyman lines to Lyman $\alpha$. Neglecting the dispersion of the redshifts, the absorption frequencies of the lines are the Lyman frequencies shifted by $b*Z_{\beta,\alpha}*c*Z_{\gamma,\alpha}*...$ where $b, c, ...$ are positive integers

This formula may be simplified by a computation of approximate values of $Z_{(\beta,\alpha)}$ and $Z_{(\gamma,\alpha)} $:

$Z_{(\beta,\alpha)} = (\nu_\beta-\nu_\alpha)/\nu_\alpha = [(1-1/3^2) -(1-1/2^2)]/(1-1/2^2) ] \approx 5/27 \approx 0,1852 \approx 3*0,0617$
 	
$Z_{(\gamma,\alpha)} = (\nu_\gamma-\nu_\alpha)/\nu_\alpha = [(1-1/4^2) -(1-1/2^2)]/(1-1/2^2) ] = 1/4 = 0,25 = 4*0,0625;$ 

In the last formula, taking into account that the gamma line is weaker than the beta, we can approximate 0,0625 by 0,0617. Thus,
 the shifts are approximately $(3b+4c)*0.0617$. 0.0617 is very close to the experimental value of the Karlsson's constant 0.061

Some values of $n = 3b+4c$ correspond to superposed, therefore intense lines: $n = 10= 3b+4c  = 3 +3 +4 = 3 +4 +3 = 4 +3 +3 $ .

\subsection{Frequency shifts of spectra emitted inside stars.}\label{gli}
Compression due to the huge surface gravity of a star similar to the Sun, leads at a short distance below the surface, despite the high temperature, the hydrogen to a quasi-crystalline state. Fixed atoms radiate sharp lines, so CREIL  is possible.

Relatively heavy atoms can be at a temperature high enough to emit X-UV rays which are redshifted while they propagate to the surface. Light emitted by lighter atoms at a lower depth and frequency, may exchange more energy with the radiation of heavier atoms than with thermal radiation, so that their frequency is increased. Emissions near the surface are not affected by CREIL.

Frequency shifts are more intense at the limb than at the center of the disk because the path in ``crystallized''  hydrogen atoms is larger.

\section{Plausible astrophysics.}
\subsection{Brilliant circles}

\begin{figure*}
\begin{center}
\resizebox{\columnwidth}{!}{\includegraphics{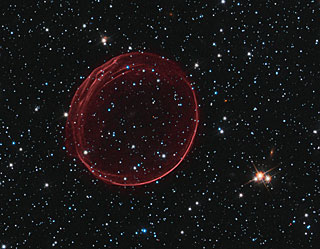}}
\end{center}
\caption{heic 1018a (Hubble).}
\label{h18}
\end{figure*}

The Hubble picture HEIC  1018A (Fig. \ref{h18}) shows an almost perfect circle that can be the image of the limb of a Str\"omgren sphere.

\subsection{Rings of supernova 1987A. }
\begin{figure*}

\begin{center}
\resizebox{\columnwidth}{!}{\includegraphics{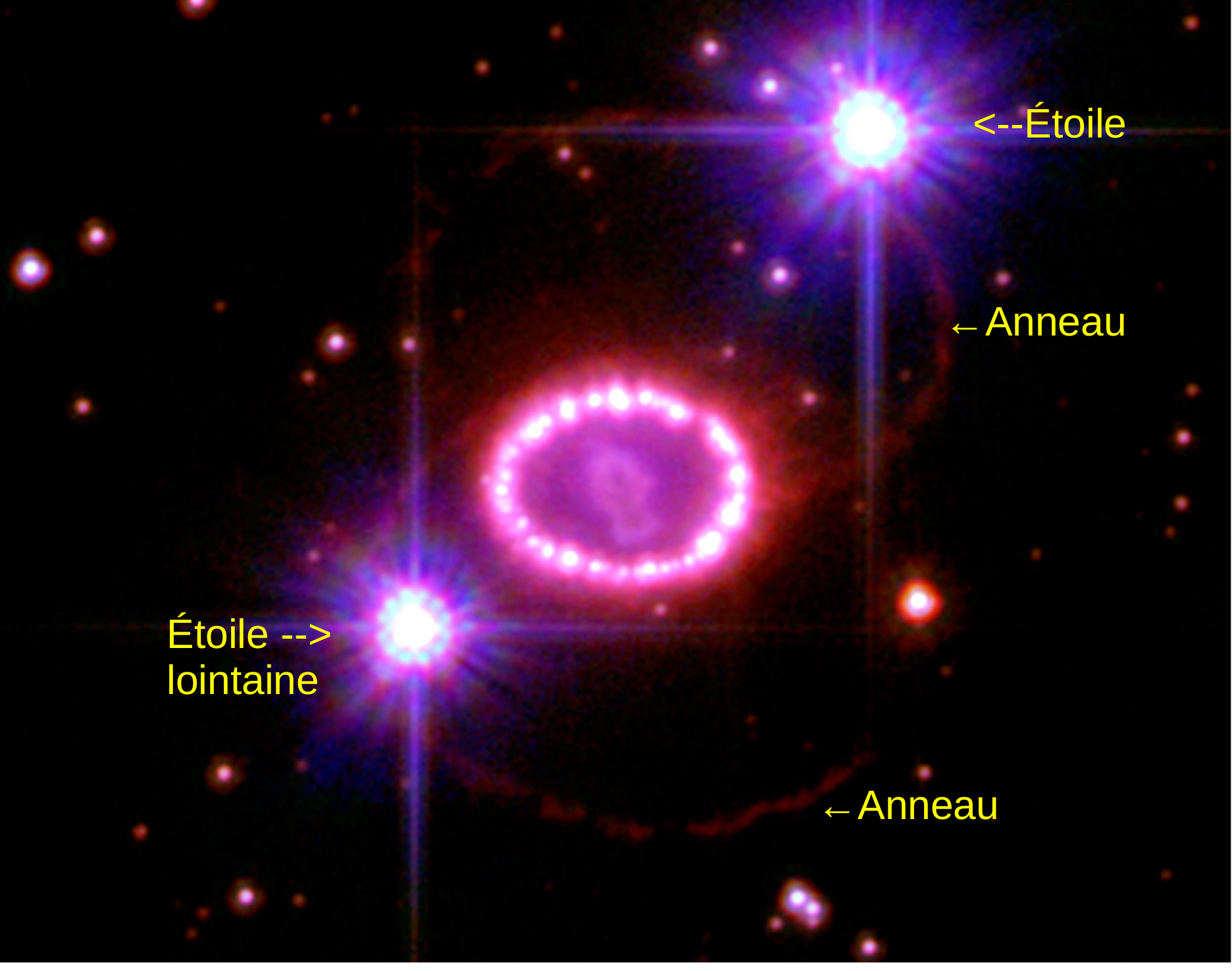}}
\end{center}
\caption{SNR1987A heic 0704a (ESA/Hubble).}
\label{snr}
\end{figure*}
If the star which creates a Str\"omgren sphere has satellites or a ring in its equatorial plane, the radiation is partially absorbed, so the cooling of the plasma is increased in this plane, the sphere is choked and takes the form an hourglass.

When the star whose name became SN1987A was lit, the astronomers observed light echoes reflected by clouds of hydrogen. The variation of the brightness of the star was used to compare the propagation time of the direct light and reflected light, so that the geometry of these clouds has been drawn in the shape of an hourglass.

Then came three rings (Image heic0704a from Hubble, Fig. \ref{snr}) at the same time as the star disappeared.

\begin{figure*}
\begin{center}
\resizebox{\columnwidth}{!}{\includegraphics{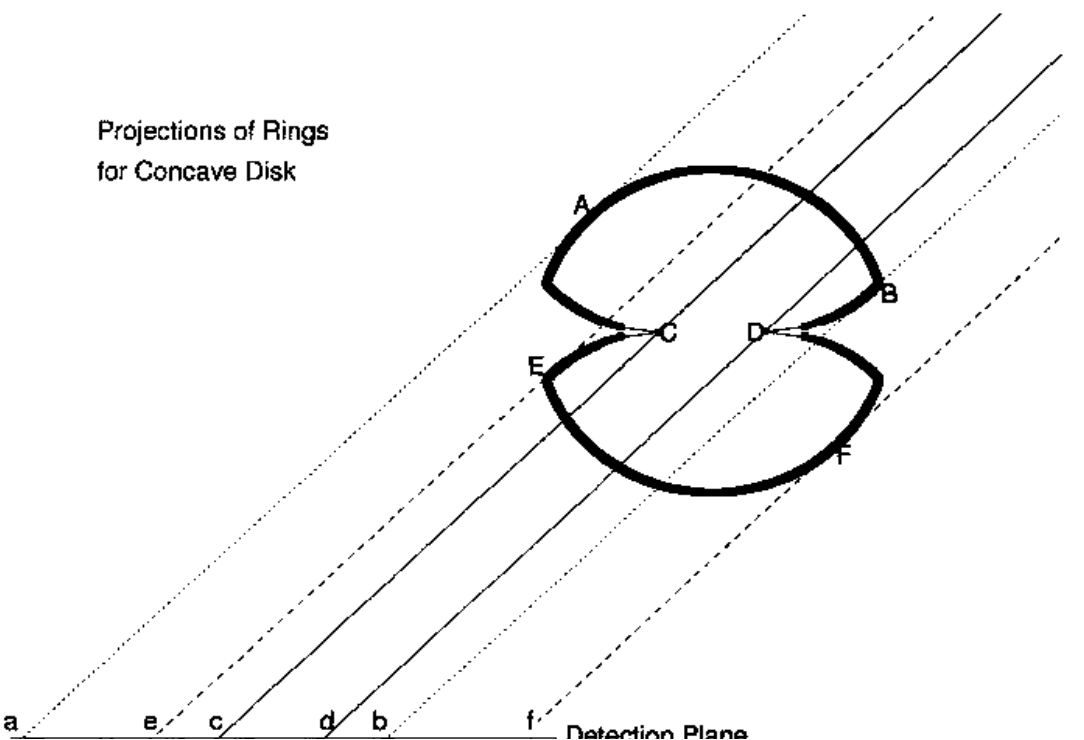}}
\end{center}
\caption{The rings of SNR1987A are the limbs of the hourglas. From Martin \& Arnett  \cite{Martin}.}
\label{sab}
\end{figure*}

Martin and Arnett \cite{Martin} have drawn a rough cut of the hourglass (Figure \ref{sab}) to show that lines tangent to the hourglass, directed towards the Earth, draw diameters of the rings. These authors were criticized by Burrows et al. \cite{Burrows}  who, disregarding the superradiance, got an image like figure \ref{sphere}, rather than figure \ref{modes}.
Coherent spectroscopy explains the ring structure, on the other hand their brightness at the expense of that of the star.

Many systems show sets of points interpreted by gravitational lensing, eg Einstein Cross. This interpretation has two problems:

- The number of alignments of heavy objects seems too large.

- The even parity of the number of points (center point excluded) in clean systems is not explained.

\subsection{Lineshapes of spectra emitted inside an hydrogen plasma, seen, for instance, inside the smallest ring of SNR1987A. }

\begin{figure*}
\begin{center}
\resizebox{\columnwidth}{!}{\includegraphics{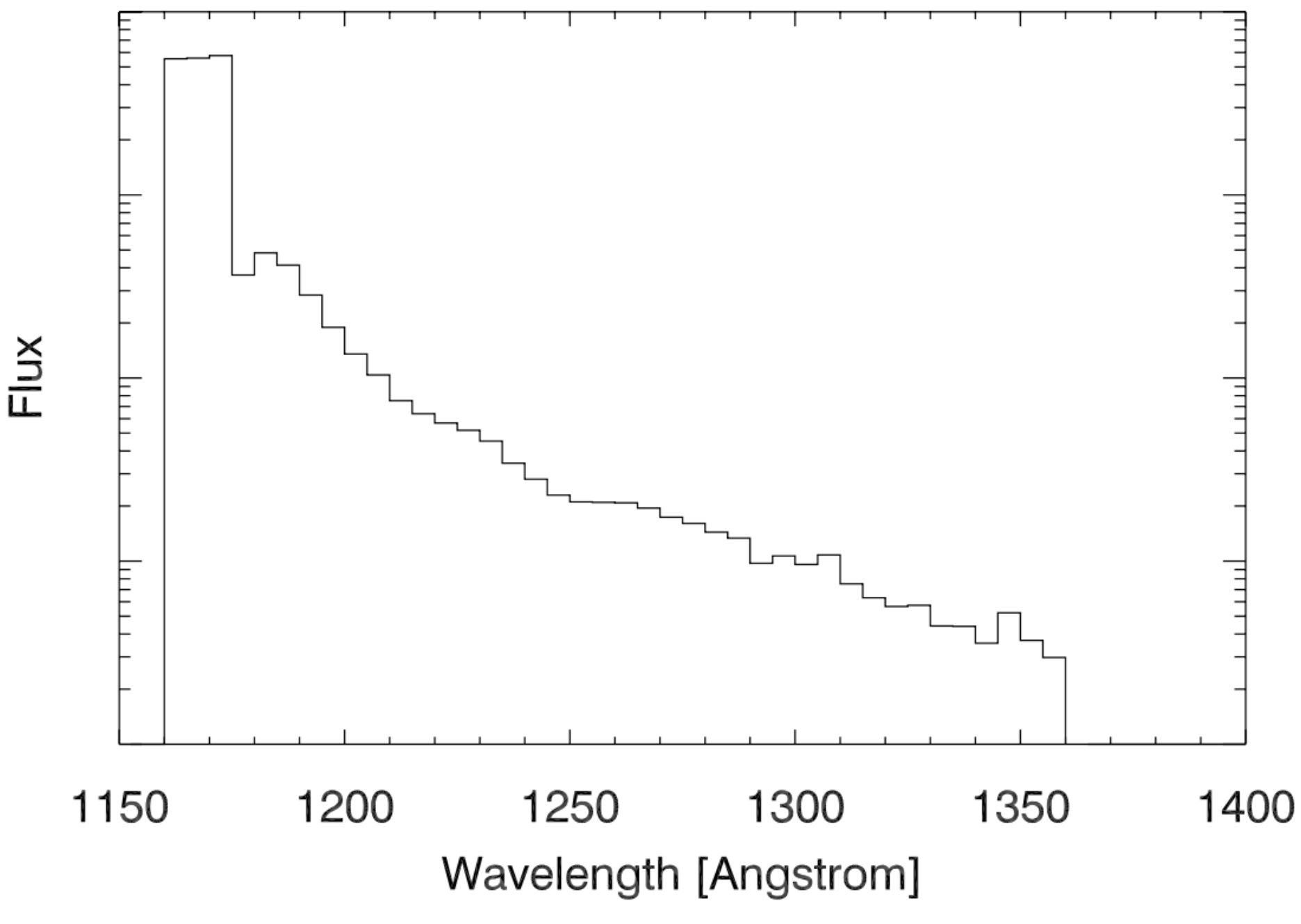}}
\end{center}
\caption{Spectrum computed by Michael et al. using a Monte-Carlo program. \cite{Michael}.}
\label{micalc}
\end{figure*}

\begin{figure*}
\begin{center}
\resizebox{\columnwidth}{!}{\includegraphics{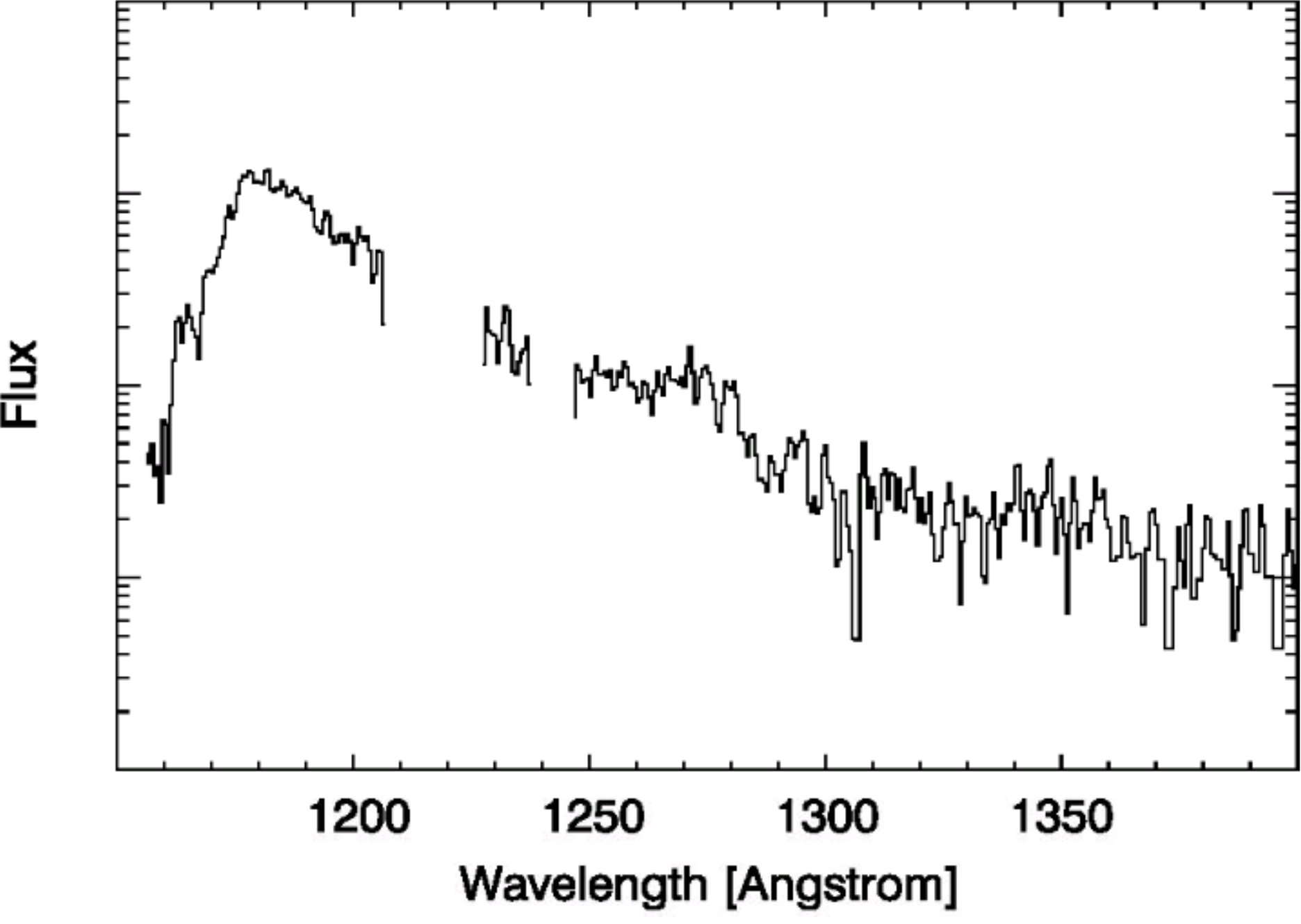}}
\end{center}
\caption{Experimental spectrum inside the small ring of SNR1987A by Michael et al. \cite{Michael}}
\label{mimes}
\end{figure*}

The spectrometer adds intensities of rays emerging from the sphere under various incidences. These rays have slightly different spectra, so that the experimental maximum (fig. \ref{mimes}) is less sharp than on figure \ref{specly}). This typical lineshape is often observed for lines emitted  by various atoms in an hydrogen plasma.

The maximum intensity in figure \ref{specly} is limited because the exponential is cut by the exit of the ray from the sphere. Instead, the calculation of Monte-Carlo (fig. \ref{micalc}) shows a maximum intensity limited by the saturation of the curve tracer.

\subsection {Spectroscopic explanation of Hubble's law.}
Assume that the density of excited atomic hydrogen H* and that the electromagnetic background irradiances are constant in the low pressure regions of universe. Neglecting the dispersion of CREIL, the redshift of the light is proportional to the column density of H*: {\it Hubble's law measures with a good approximation the column density of n=2 atomic hydrogen}. In cold regions of the nebulae, the density of H* seems constant, so that Hubble's law measures a distance. But the density of H* may be large close to hot objects. 

Many astronomers, Halton Arp in particular \cite{Arp}, remarked that sets of stars which seem bound may have different redshifts, the hottest being the most redshifted. As hydrogen plasma often contain a large density of H*, some people try to explain their power to redshift light by recoil effects on free protons or electrons, but there is no hope to find such an effect in an homogeneous plasma.

\subsection{Intergalactic voids.}
Some galaxies contain massive and hot stars emitting  hot hydrogen winds. Some winds remain excited about until some other galaxies. The bubbles of  H*  thus formed are swollen on maps by a blind use of Hubble's law. {\it The radial distances of galaxies maps and transverse distances that are deducted are exaggerated.} This results in a miscalculation of the dynamics of spiral galaxies, corrected by the improbable hypothesis of dark matter.

From a local perspective, the bubbles increase irregularly radial distances, generating voids which contain few stars, creating an observed spongy structure in the maps of galaxies. 

\subsection{Experimental observation of CREIL.}
\begin{figure*}
\begin{center}
\resizebox{\columnwidth}{!}{\includegraphics{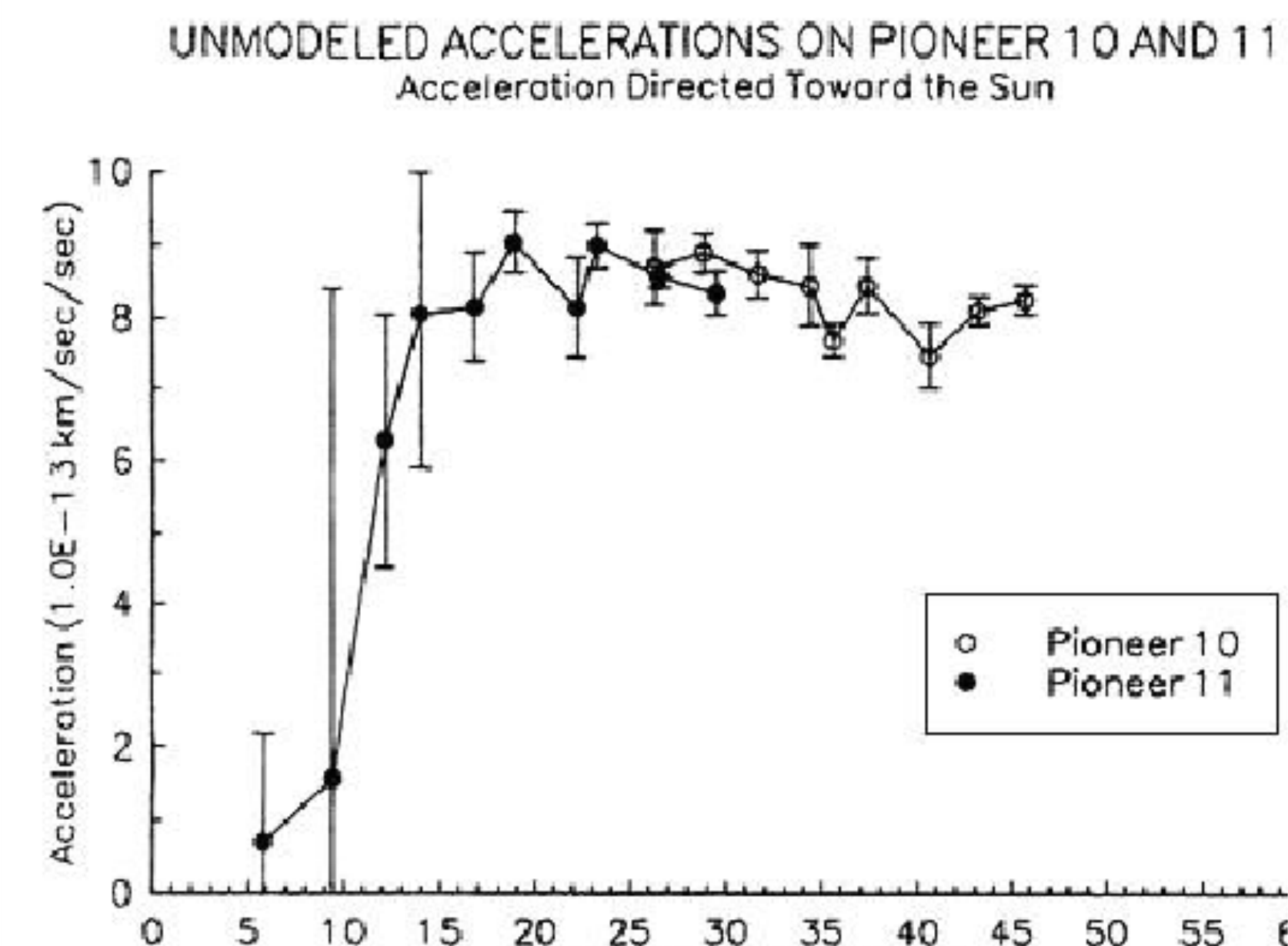}}
\end{center}
\caption{Pioneer 10 and 11 “ anomalous acceleration'' . From Anderson et al \cite{Anderson}}
\label{pioneer}
\end{figure*}
This check requires a long light path in H* at low pressure. A rough calculation of adiabatic expansion of hydrogen plasma ($p^+ + e^- $) making the solar wind produces  H* at 10 AU, then de-excites it at 15 AU. It is a weak type of Str\"omgren shell.


Sunlight is redshifted as it passes through this shell, producing a pulsed excitation of the spin-coupling states of H* atoms, excitation  detected by a continuous wave probe beam. 

The carrier of the signals exchanged with the Pioneer 10 and 11 probes undergoes an increase in frequency during its passage through the layer containing H* (Fig. \ref{pioneer}). Note that this frequency increases during the pulses of the sunlight, with the destruction of the temporal coherence while the frequency changes. The detector extracts a continuous wave from the noisy one it receives. Its frequency is increased.

This interpretation could be tested by not measuring the frequency of a carrier, but the frequency of  pulses which would be affected by a Doppler effect, but not by a  CREIL. The experiment may be very long because the frequency shift is almost constant beyond 15AU.
Thermal radiation observed at 2.7 K can be amplified by the same mechanism. Its anisotropy observed related to the ecliptic, is explained by the anisotropy of the solar wind.

\subsection{Periodicities of the redshifts of quasars and galaxies. }
The theory in \ref{peri} explains the periodicities observed in the spectra of the quasars: the propagation of light is divided into cycles which during the one hand redshift light, and, on the other hand, whenever an absorbed line gets the Lyman $\alpha$ frequency, inscribe emission and absorption lines of the gas. These cycles stop where the local gas is not anymore able to produce low pressure 2P hydrogen, at any time of a cycle. Considering a large set of observed objects, the probability to observe an object whose final redshift corresponds to a stop by coincidence of an absorbed line with the Ly $\alpha$ line is large. 

In 1968, Burbidge \cite{Burbidge} then many other authors found experimentally this result: The redshifts of the galaxies are often given by $ Z_{obs}\approx 0.061 n $ where $n$ has certain integer values, larger than 2, and $0.061$ is the Karlsson periodicity. 
\subsection{SOHO observation of UV-X  spectrum of the Sun }
\begin{figure*}
\begin{center}
\resizebox{\columnwidth}{!}{\includegraphics{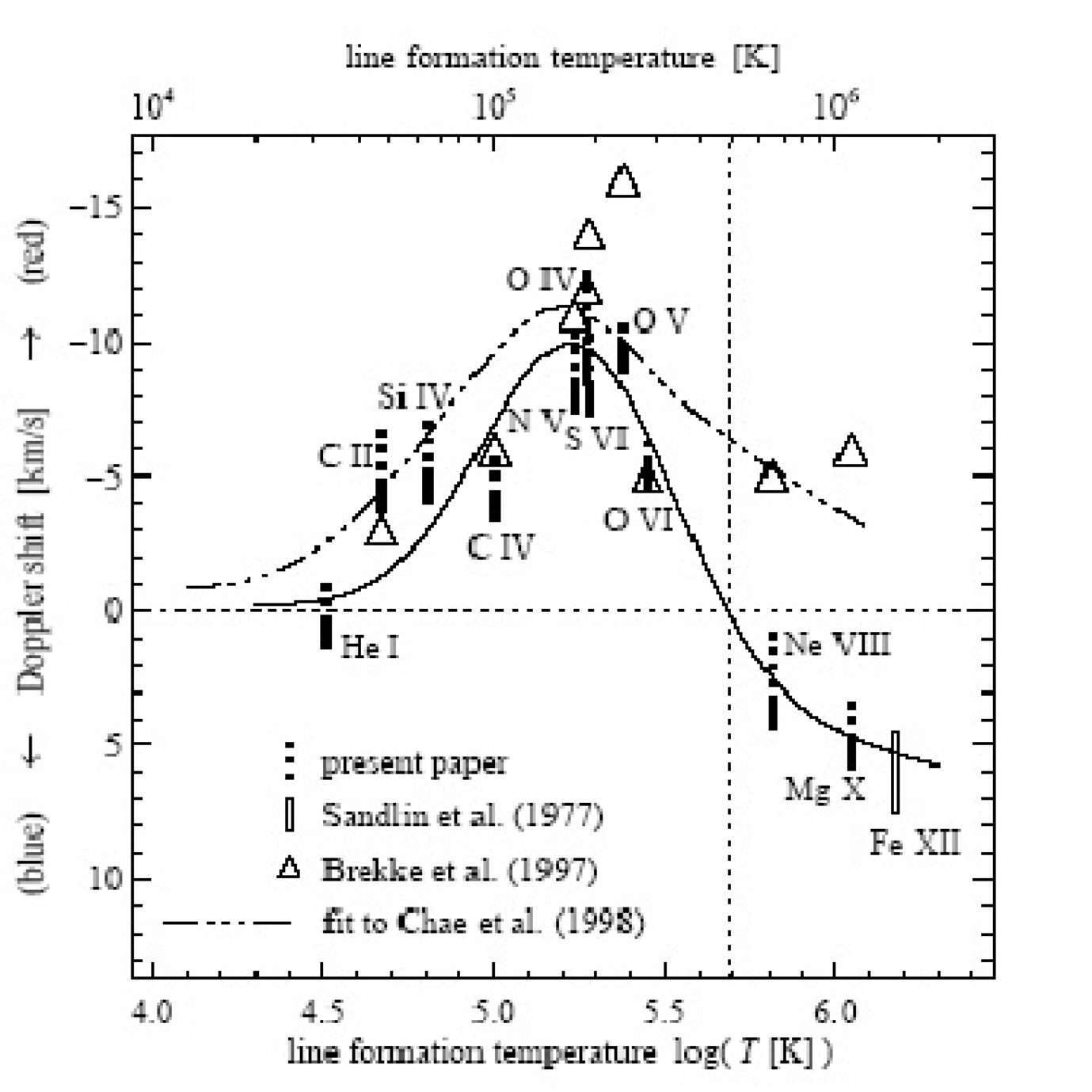}}
\end{center}
\caption{UV-X spectrum of the Sun \cite{Anderson}.}
\label{soho}
\end{figure*}

Peter \& Judge \cite{Peter} attribute the frequency shifts of X-UV rays of Sun to Doppler effect produced by a vertical velocity of the emitting gas. With this hypothesis, the maximum effect is at the center of the disc, and it is null at the limb whose radiation is used to recalibrate the spectrometer. But, using the internal calibration, the frequencies observed at the limb are further from laboratory frequencies than those observed at the center.

With the theory given in \ref{gli}, the meaning of the y-axis of the original figure \ref{soho} and scale must be changed because this figure represents the difference between the frequency shift (minimum) at the center of the disc and the frequency  shift (maximum) at the limb. Thus, the internal calibration appears good. 

\section{Conclusion.}
Most physicists have long held that the coherent interactions of light and matter are negligible, with the exception of refraction. Little affected by the discovery of the laser, astrophysicists have kept this vision of coherent interactions expressed by Menzel in 1931 and confirmed by an aberrant concept of the photon as a small particle able to interact only with a single atom in a gas at low pressure. This conception, incompatible with quantum electrodynamics, has been strongly criticized by W. E. Lamb jr., W. P. Schleich, M. O. Scully and C. H. Townes \cite{WLamb2}.
 
\medskip
Some astrophysicists have sought to interpret the redshift of the spectra of distant stars by a gradual shift of the frequency of a sine wave propagating in a suitable medium. These attempts were doomed to failure because the introduction of a local variation in frequency into a sine wave introduces one or more discontinuities.

Frequency variations are observed in short pulses from femtosecond lasers. It is ``Impulsive Stimulated Raman Scattering`` (ISRS). These variations are approximately inversely proportional to the cube of the pulse duration. Replacement of femtosecond pulses by pulses $10^{5}$ times longer that model natural incoherent light, reduces the frequency shift by a factor $10^{15}$, so that an observation requires an astronomical path.

This ISRS in incoherent natural  light requires a gas having a quadrupole resonance frequency below 1 GHz, but as high as possible. Atomic hydrogen in excited levels 2S or 2P (denoted H*) is suitable. It is found in hydrogen plasma, and 2P H* is formed in cold atomic hydrogen by  Lyman $\alpha$ optical pumping. In practice, several effects must be coupled into a {\it parametric effect} named  Coherent Raman Effect on Incoherent Light (CREIL) so that the Raman levels are not excited. The CREIL exchanges progressively energy between beams of light, in accordance with thermodynamics.

\medskip
The use of optical coherence allows to build models that appear to explain many observations, in particular:

- A bright star with a ring or satellites and immersed in a cloud of hydrogen  has the appearance, the hydrogen spectrum and the evolution of SNR1987A.

 - It is surprising that not a single ``accreting neutron star''  has been seen in close hydrogen clouds, when they should be bright. In fact many have been observed, but they are called ``quasars''.

 - The anomalous acceleration of Pioneer 10 and 11 probes is a consequence of the blueshift of the microwave carrier of the signals exchanged between the probe and Earth : The adiabatic expansion of the solar wind creates H* between 10 and 15 AU, which catalyzes a transfer of energy from incoherent sunlight to microwaves.

 - The strong reddening of light in the vicinity of hot stars that produce a large amount of H*, are usually interpreted as voids, so that the maps of galaxies are spongy.

\medskip
Our models must be improved, introducing gas other than hydrogen, making them more quantitative; other models may be easily set to study other observations: There remains a lot of work to develop our start of interpretation of astronomical observations using optical coherence.

\medskip
Optical coherence could be an alternative to the Big Bang theory. But many years seem necessary for the adoption of the upheaval that is the substitution of a basic, experimentally verified theory, to a wonderful theory.

\section{Appendix: Frequency shifts of time-incoherent light beams by coherent transfers of energy (ISRS or CREIL).}
                                                                                                                                                                                                                    
Let us suppose that each element of volume considered by Huygens contains molecules which emit also a wavelet of much lower amplitude at same frequency, whose phase is delayed by $\pi/2$ (Rayleigh coherent emission). Both types of emission generate the same wave surfaces, so that their emitted fields may{} be simply added. Are $E_0 \sin (\Omega T)$ the\label{shell} incident field, $E_0K\epsilon \cos(\Omega T)$ the field diffused in a layer of infinitesimal thickness $\epsilon=c\mathrm dt$ on a wave surface, and $K$ a coefficient of diffusion. The total field is:

\begin{equation}
E=E_0[\sin(\Omega t)+K\epsilon \cos(\Omega t)]\label{refr}
\end{equation}

\begin{eqnarray}
\approx E_0[\sin(\Omega t)\cos(K\epsilon)+\sin(K\epsilon )\cos(\Omega t)]=\nonumber\\
= E_0\sin(\Omega t -K\epsilon).
\end{eqnarray}
                                                                                                                                                                                                                     
 This result defines the index of refraction $n$ by the identification:
\begin{equation}
K=2\pi n/\lambda=\Omega n/c.\label{index}
\end{equation}

Add, in this theory of refraction, to the Rayleigh scattering, a Raman scattering, with a shifting frequency $\omega$, but no initial phase shift. Neglecting refraction to simplify formulas and setting $K'>0$ the anti-Stokes diffusion coefficient, equation \ref{refr} becomes:
\begin{equation}  
E=E_0[(1-K'\epsilon)\sin(\Omega t)+K'\epsilon \sin((\Omega+\omega)t)].
\end{equation}

In this equation, incident amplitude is reduced to obtain the balance of energy for $\omega=0$.

\begin{eqnarray}
E=E_0\{(1-K'\epsilon)\sin(\Omega t)+\nonumber\\
+K'\epsilon[\sin(\Omega t)\cos(\omega t)+\sin(\omega t)\cos(\Omega t)]\}.
\end{eqnarray}

$K'\epsilon$ is infinitesimal; suppose that between the beginning of a pulse at $t=0$ and its end, $\omega t$ is small; the second term cancels with the third, and the last one transforms:

\begin{eqnarray}
E\approx E_0[\sin\Omega t+\nonumber\\+\sin(K'\epsilon\omega t)\cos(\Omega t)]\\
E\approx E_0[\sin(\Omega t)\cos(K'\epsilon\omega t)+\nonumber\\+ \sin(K'\epsilon\omega t)\cos(\Omega t)\\
E\approx E_0\sin[(\Omega+K'\epsilon\omega)t].\label{eq4}
\end{eqnarray}

Hypothesis $\omega t$ small requires that Raman period $2\pi/\omega$ is large in comparison with the duration of the light pulses; to avoid large perturbations by collisions, the collisional time must be larger than this duration.  This is a particular case of the condition of space coherence and constructive interference written by G. L. Lamb \cite{GLamb}.

Stokes contribution, obtained replacing $K'$ by a negative $K''$, must be added. Assuming that the gas is at a high temperature $T$, $K'+K''$ is proportional to the difference of populations in Raman levels, that is to $\exp[-h\omega/(2\pi kT)]-1 \propto \omega/T$.

$K'$ and $K''$ obey a relation similar to relation \ref{index}, where Raman polarisability which replaces the index of refraction is also proportional to the pressure of the gas $P$ and does not depend much on the frequency if the atoms are far from resonances; thus, $K'$ and $K''$ are proportional to $P\Omega$, and $(K'+K'')$ to $P\Omega \omega/T$. Therefore, for a given gas, the frequency shift is:
\begin{equation}
\Delta\Omega=(K'+K'')\epsilon\omega\propto P\epsilon\Omega\omega^2/T.\label{delom}
\end{equation}
The relative frequency shift $\Delta\Omega/\Omega$ of this space-Coherent Raman Effect on time-Incoherent Light (CREIL) is nearly independent on $\Omega$ and proportional to the integral of $Pc{\rm d}t$, that is to the column density of active gas along the path.

The path needed for a given (observable) redshift is inversely proportional to $P\omega^2$. At a given temperature, assuming that the polarisability does not depend on the frequency, and that $P$ and $\omega$ may be chosen as large as allowed by Lamb's condition, $P$ and $\omega$ are inversely proportional to the length of the pulses. Thus, the path needed for an observation is inversely proportional to the cube of the length of the pulses: an observation easy in a laboratory with 10 femtosecond pulses requires 
$10^{15}$ times longer, astronomical paths with nanosecond pulses making ordinary incoherent light. 

\end{multicols}

\end{document}